\documentclass[aps,prl,twocolumn,superscriptaddress,amsmath,amssymb,nofootinbib,floatfix,10pt]{revtex4-2}
\usepackage{graphicx}
\usepackage{bm}
\usepackage[colorlinks=true,linkcolor=blue,citecolor=blue,urlcolor=blue]{hyperref}
\usepackage{xcolor}
\usepackage{orcidlink}

\newcommand{\Msun}{M_\odot}
\newcommand{\gcc}{\,{\rm g\,cm^{-3}}}
\newcommand{\ms}{\,{\rm ms}}
\newcommand{\us}{\,\mu{\rm s}}
\newcommand{\GRoneD}{\texttt{GR1D}}

\begin{document}
	
	\title{Quark deconfinement as a natural trigger of Type-I critical collapse:\\black holes from neutron stars deemed stable}
	
	\author{Hrishikesh Borthakur\,\orcidlink{0009-0003-4580-1370}}
	\email{hrishikesh24@iiserb.ac.in}
	\affiliation{Department of Physics, Indian Institute of Science Education and Research Bhopal, Bhopal 462066, India}
	
	\author{Ritam Mallick\,\orcidlink{0000-0003-2943-6388}}
	\email{mallick@iiserb.ac.in}
	\affiliation{Department of Physics, Indian Institute of Science Education and Research Bhopal, Bhopal 462066, India}
	
	\date{\today}
	
	\begin{abstract}
Spherically symmetric general-relativistic simulations show that a first-order hadron--quark phase transition can collapse radially stable neutron stars into black holes, even when a stable hybrid star of the same baryon mass exists. The dynamical threshold lies $0.12\,M_\odot$ in baryon mass below the static one, so black holes can be lighter than the heaviest stable star. The threshold is a Type-I critical point: near it, the star lingers at an unstable hybrid star of equal baryon mass for a time set by that star's unstable mode. The effect persists within a narrower window for finite conversion rates.
	\end{abstract}

	\maketitle
	
	\textit{Introduction.}---How light can a black hole (BH) born from a neutron star (NS) be? The usual answer is fixed by statics: a star collapses only when no stable configuration of its baryon mass exists, so such BHs should be heavier than the most massive stable star. A first-order hadron--quark phase transition (PT) puts this picture to a sharp test, because the conversion of a metastable hadronic core is sudden and violent~\cite{Bombaci2004,Lin2006,Abdikamalov2009}. Here we show that statics can fail: a radially stable NS can collapse on conversion even when a radially stable hybrid star of the same baryon mass exists, leaving a BH lighter than the heaviest stable hybrid star. The boundary between collapse and survival is a critical point of the kind known from gravitational collapse ~\cite{choptuik1993,Gundlach2007}.
    
    
    
    
If such a route exists in nature, BHs would form from NSs that are radially stable by every static criterion, and the heaviest NSs and the lightest BHs formed in this way would overlap in mass. The difficulty is the trigger. Driving a stable NS near its maximum mass to collapse requires a coherent inward velocity of $\sim0.01$--$0.1\,c$ across the star~\cite{Novak2001}, i.e.\
$\sim10^{50}$--$10^{52}$~erg of bulk kinetic energy delivered within a dynamical time of $\sim0.1\ms$. Spin-down, cooling, and accretion act quasi-statically, and crustal or magnetic events release orders of magnitude less energy. A first-order PT from hadronic to deconfined quark matter does not share this limitation: conversion removes the pressure support of the core within a
dynamical time and, for the EoS used here, releases $\sim10^{52}$~erg, enough to drive the star to the threshold from within.


  If the hadron--quark transition is of first order, the mass--radius relation develops a hybrid branch whose stability is controlled by the size of the energy-density jump~\cite{Seidov1971,Alford2013,Christian2018,Benic2015,haque2026}.Because quark matter must first be nucleated, purely hadronic stars can survive as metastable configurations above the onset density~\cite{PhysRevD.77.083002,Bombaci2004,Drago2014}. Once the core converts, the star either settles on the hybrid branch or collapses to a BH~\cite{Lin2006,Abdikamalov2009,Dimmelmeier2009,prasad2018}. PTs may also leave imprints in binary NS mergers~\cite{Most2019,Bauswein2019,haque2025}.     The static criterion predicts BH formation after baryon-mass-conserving conversion only when $M_B>M_{B,\max}$ of the hybrid branch, but the sudden core-pressure drop can drive coherent infall that overshoots the same-$M_B$ hybrid equilibrium and crosses the unstable configuration, causing collapse despite an available stable endpoint. The migration–collapse boundary is therefore a dynamical threshold naturally described by critical phenomena~\cite{choptuik1993,Gundlach2007}, with Type-I behavior mediated by an unstable equilibrium star already found in linearly unstable NS models~\cite{Radice2010} and driven collapse~\cite{Noble2016}. Velocity perturbations can likewise collapse stable NSs~\cite{Novak2001,Noble2016} and trigger migration between twin-star branches, while some hybrid stars collapse instead~\cite{99c1-2p2y}.
	

In this Letter, we evolve metastable hadronic stars with an EoS containing a strong first-order PT and show that conversion naturally triggers BH formation without external perturbations, even from radially stable stars with stable hybrid counterparts of equal baryon mass $M_B$. The collapse boundary exhibits Type-I criticality, governed by the unstable hybrid star at the same $M_B$, with the near-critical lifetime set by the unstable star's fundamental radial-mode growth rate. Radially perturbed stable hybrid stars approach the same critical solution, identifying conversion as an internal trigger for velocity-induced collapse and producing BHs below the hybrid branch’s maximum mass, where static stability analysis predicts a stable remnant.
	
	
	\textit{PT-induced collapse.}---The EoS~\cite{Saha:2025} is tabulated in $(\varepsilon,P,n_B)$ and has a Maxwell transition at $P_t=99.9\,{\rm MeV\,fm^{-3}}$, from $\rho_t=7.05\times10^{14}$ to $\rho_q=1.00\times10^{15}\gcc$. The energy-density jump, $\Delta\varepsilon/\varepsilon_t=0.518$, lies below the Seidov limit $1/2+3P_t/(2\varepsilon_t)=0.852$~\cite{Seidov1971}, so the hybrid branch is connected to the hadronic one. For use in \GRoneD~\cite{OConnor2010} we fit the table with a causal piecewise polytrope~\cite{Read2009}. The fit has an SLy crust~\cite{Douchin2001}, three hadronic segments, the transition, and six quark segments, and reproduces the pressure to $0.012$~dex rms~\cite{SM}. It gives $M_{\rm TOV}=2.1568\,\Msun$, hybrid onset at $1.921\,\Msun$, $R_{1.4}=12.22$~km, and $\Lambda_{1.4}=462$, consistent with recent NICER and GW170817 measurement \cite{Fonseca2021,Salmi2024,Choudhury2024,Vinciguerra2024,Choudhury2024,Abbott2018}[Fig.~\ref{fig:eos}]. At a Maxwell transition, $c_s^2=0$, which makes the hydrodynamic system only weakly hyperbolic. We therefore replace the plateau by a mixed phase with $P\propto\rho^{\Gamma_{\rm mix}}$ and $\Gamma_{\rm mix}=0.5$, which spans $6.8\times10^{14}$ to $1.0\times10^{15}\gcc$~\cite{SM}. This barely changes the EoS but keeps the sound speed finite.
	
	We solve the Einstein--Euler system in spherical symmetry with \GRoneD\ (radial gauge, polar slicing, HLLE fluxes~\cite{Harten1983}, slope-limited second-order reconstruction~\cite{vanLeer1977}), extended for piecewise-polytropic EoSs with a $\Gamma$-law thermal part ($\Gamma_{\rm th}=1.75$), TOV initial data, and constraint-satisfying velocity perturbations~\cite{Gourgoulhon1992}. The grid is uniform with $\Delta r=20$~m out to $20$~km and is geometrically stretched beyond that, up to $500$~km. A BH is declared when the central lapse falls below $\alpha_c=0.05$.
	
\begin{figure}[t]
		\includegraphics[width=\columnwidth]{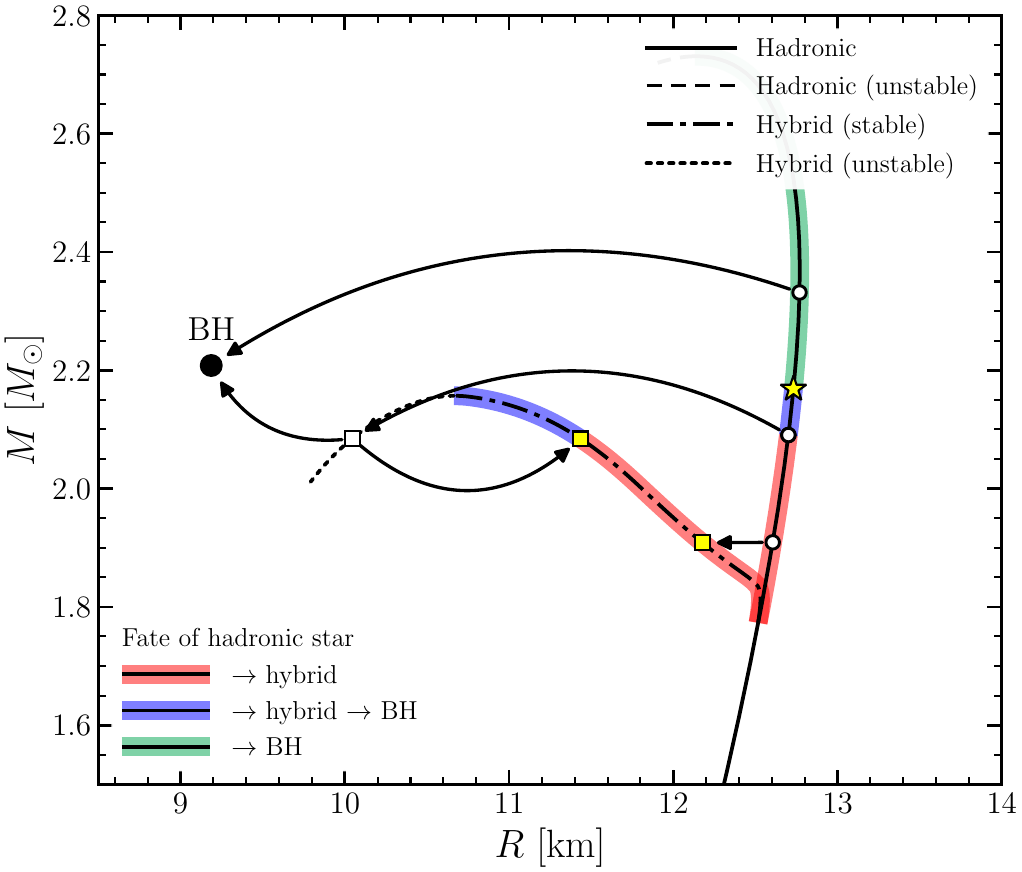}
		\caption{PT-induced collapse of cold NSs without external perturbation. All $M$--$R$ relations are drawn in black: hadronic stars as solid lines and hybrid stars as dash-dotted lines, with dotted (hybrid) lines marking unstable branches. The two relations coincide below the transition onset (unshaded). Above the onset the hadronic stars are metastable, and the shading shows the outcome of core conversion at fixed $M_B$. Red marks migration to the stable hybrid star with the same $M_B$. Blue marks dynamical collapse to a BH even though such a hybrid star exists. Green marks collapse because no hybrid counterpart exists, $M_B>M_{B,\max}$. The same red and blue shading on the stable hybrid branch marks the hybrid stars with the $M_B$ of the corresponding hadronic segment. The yellow star marks the hadronic configuration with $M_B=M_{B,\max}$. Arrows follow representative simulated stars (open circles) to their end states, which are stable hybrid stars (filled squares) or a BH. The open square is the unstable hybrid star with the threshold $M_B$, i.e.\ the critical solution. The boundary between the red and blue segments is the dynamical threshold, the critical point that separates subcritical from supercritical evolutions.}
		\label{fig:pt}
	\end{figure}

	We take metastable hadronic TOV stars and evolve them with the hybrid EoS, without any perturbation, so that the core converts at $t=0$. Figure~\ref{fig:pt} summarizes the outcomes. The static threshold, above which no hybrid star has the same $M_B$, lies at $M=2.1684\,\Msun$ ($M_B=2.5581\,\Msun$); stars above it collapse within $\approx0.3$~ms. Stars from the onset at $1.921\,\Msun$ up to $2.086\,\Msun$ convert, bounce, and oscillate about the stable hybrid star with the same $M_B$. In between lies a band of \emph{dynamical} collapse: stars with $2.086\lesssim M\le2.1684\,\Msun$ form BHs although a stable hybrid star with the same $M_B$ exists. Bisection in the initial central density $\rho_{c,0}$ locates the dynamical threshold at $\rho_{c,0}^{\star}=7.3729435\times10^{14}\gcc$, i.e.\ $M^{\rm dyn}=2.086\,\Msun$ ($M_B=2.441\,\Msun$), $0.117\,\Msun$ in baryon mass below the static threshold. Repeating the bisection with $\Delta r=40$~m shifts $\rho_{c,0}^{\star}$ by $1.6\times10^{-4}$ and $M^{\rm dyn}$ by $\sim5\times10^{-5}\,\Msun$.

	In spherical symmetry, no gravitational waves are emitted, and the total
	gravitational mass on the grid is conserved by the evolution: in a
	representative collapse run, evolved until shortly before horizon formation,
	it changes by $8.0\times10^{-5}\,\Msun$ (a relative drift of $3.5\times10^{-5}$). The mass of the BH can therefore not exceed that of the initial star. Across the band, the initial masses range from
	$M^{\rm dyn}=2.086\,\Msun$ to $2.1684\,\Msun$, so all but the
	uppermost $0.012\,\Msun$ of the band lie below the maximum mass $M_{\rm TOV}=2.1568\,\Msun$ of the hybrid branch. Core conversion thus forms BHs lighter than the most massive stable star of the same EoS, and the masses of BHs and stable hybrid stars produced by this channel overlap.
	
	To test whether the overshoot depends on the transition strength, and to make sure it is not an artifact of a particular equation of state, we construct a second set of hybrid EoSs in which the transition density is fixed (onset at $M\simeq1.6\,M_\odot$) and only the density jump $\Delta\rho/\rho_t$ is varied (Table~\ref{tab:thresholds}).
	For every member of this set, radially stable hadronic stars below the static
	threshold still collapse, and near the dynamical threshold $\rho_{c,0}^*$ the
	evolution again approaches the unstable hybrid star of equal baryon mass, with
	lifetimes following the Type-I scaling. The window $\Delta M_B$ below the static limit
	decreases with increasing transition strength, from $0.075$ to $0.052\,M_\odot$.
	Because $M_{B,\rm TOV}$ falls faster than the window shrinks, however, the absolute
	threshold $M_{B,*}$ decreases, and the range of hadronic stars that collapse widens.
	See Sec.~\ref{sec:Strength_PT} of the Supplemental Material for details.

	\begin{table}[t]
		\caption{Static and dynamical collapse thresholds for the hybrid EoSs
			(onset at $M\simeq1.6\,M_\odot$, $\Gamma_{\rm PT}=0.5$).
			$M_{\rm TOV}$, $M_{B,\rm TOV}$: maximum gravitational and baryon mass of stable hybrid stars;
			$M_*$, $M_{B,*}$: gravitational and baryon mass of the hadronic initial star at the
			dynamical threshold; $\Delta M_B = M_{B,\rm TOV}-M_{B,*}$ is the width of the
			collapse window, and $\Delta M = M_{\rm TOV}-M_*$ the amount by which the lightest BH
			formed by conversion lies below $M_{\rm TOV}$. Masses in $M_\odot$.}
		\label{tab:thresholds}
		\begin{ruledtabular}
			\begin{tabular}{cccccc}
				$\Delta\rho/\rho_t$ & $M_{\rm TOV}$ & $M_{B,\rm TOV}$ & $M_*$ & $\Delta M$ & $\Delta M_B$ \\
				\hline
				hadronic & 2.326 & 2.749 & --    & --     & --     \\
				0.15     & 2.091 & 2.427 & 2.039 & 0.0512 & 0.0748 \\
				0.30     & 1.969 & 2.263 & 1.920 & 0.0484 & 0.0691 \\
				0.45     & 1.866 & 2.126 & 1.829 & 0.0373 & 0.0524 \\
			\end{tabular}
		\end{ruledtabular}
	\end{table}
	
\begin{figure}[t]
		\centering
		\includegraphics[width=\columnwidth]{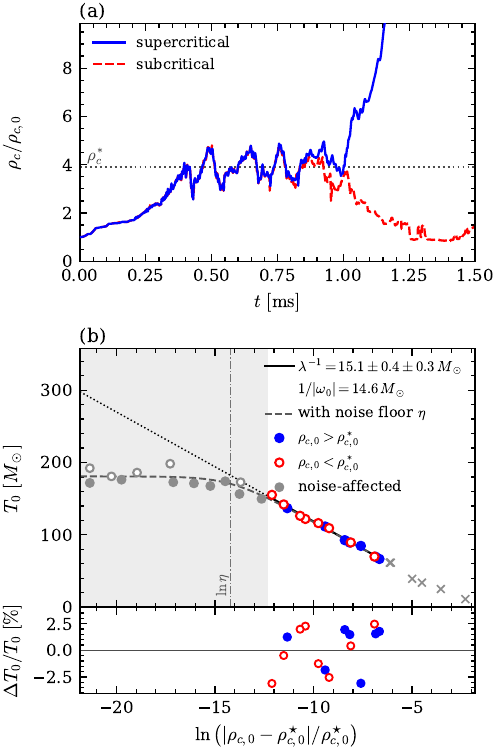}
		\caption{Critical behavior of PT-induced collapse.
			(a) Central density, normalized to its initial value, for the two runs bracketing the threshold, $\rho_{c,0}=7.372943468730\times10^{14}$ (solid, collapse) and $7.372943464799\times10^{14}\gcc$ (dashed, no collapse). The dotted line marks $\rho_c^{*}$ of the critical hybrid star. A 21-point running median removes isolated numerical spikes (display only).
			(b) Lifetime $T_0$ against the distance from the threshold for $\epsilon=0.05$. Coloured symbols mark the scaling range used in the fit (filled: collapse, open: no collapse). Grey circles mark noise-affected runs close to the threshold (shaded), and crosses mark pre-asymptotic runs. The solid line has the $\epsilon$-averaged slope of Eq.~\eqref{eq:lambda_pt}, and the dotted line extrapolates it. The dashed curve is Eq.~\eqref{eq:T0_noise}, with $\ln\eta$ marked by the dash-dotted line. The lower panel shows the relative residuals in the scaling range.}
		\label{fig:ptcrit}
	\end{figure}

	\textit{Critical behavior.}---Close to the threshold, the converting star contracts to a hybrid configuration with a large quark core and hovers there before collapsing or rebounding [Fig.~\ref{fig:ptcrit}(a)]. During the hover the time-averaged central lapse is $\alpha_c=0.241$, and the time-averaged central density is $\rho_c^{*}=(2.87\pm0.04)\times10^{15}\gcc$. The unstable hybrid star with the same $M_B=2.441\,\Msun$ has $\alpha_c=0.240$, $M=2.086\,\Msun$, and $R=10.05$~km. This is the intermediate attractor of Type-I critical collapse~\cite{Gundlach2007,Radice2010,Noble2016}. Near threshold, the time $T_0$ spent near the attractor then scales as
	\begin{equation}
		T_0 = C - \lambda^{-1}\ln|p-p^{*}| ,
		\label{eq:T0_scaling}
	\end{equation}
	where $p$ parametrizes the initial data and $\lambda$ is the growth rate of the attractor's single unstable mode. We measure $T_0$ relative to the marginally subcritical run~\cite{Noble2016}: it is the time at which the central density, averaged over one pulsation period, departs permanently from that run by a fraction $\epsilon$. All times are measured in \GRoneD\ coordinate time, corresponding to time at infinity. Details are given in Sec.~\ref{sec:SM_lifetimes} of the Supplemental Material~\cite{SM}. With $p=\rho_{c,0}$ and the dimensionless distance $|\rho_{c,0}-\rho_{c,0}^{\star}|/\rho_{c,0}^{\star}$, collapsing and non-collapsing runs lie on one line [Fig.~\ref{fig:ptcrit}(b)]. Averaging over $\epsilon=0.02$, $0.05$, and $0.10$ gives
	\begin{equation}
		\lambda^{-1}=15.1\pm0.4\,({\rm stat})\pm0.3\,({\rm sys})\,\Msun=74\pm3\us ,
		\label{eq:lambda_pt}
	\end{equation}
	and $14.4\pm0.9\pm0.3\,\Msun$ at $\Delta r=40$~m. The collapse times of the supercritical runs, which require no lifetime extraction, give $14.3\pm0.1\,\Msun$. Solving the radial-pulsation eigenvalue problem~\cite{chandrasekhar1964,Kokkotas2001} for the unstable hybrid star yields an unstable fundamental mode with
$1/|\omega_0|=71.9\us=14.6\pm0.2\,\Msun$, where the error reflects the
uncertainty of $\rho_c^{*}$.
	
	Below a distance of $\sim10^{-6}$ from the threshold, the lifetimes stop growing
	and scatter about a common value [shaded region of Fig.~\ref{fig:ptcrit}(b)];
	the collapse times recorded during the bisection behave in the same way. We
	attribute this floor to the onset of conversion. As the center enters the mixed
	phase, every run receives a small perturbation that is set by the
	discretization rather than by its distance from the threshold, and once that
	distance becomes comparable to the perturbation, the lifetime no longer grows.
	An effective offset $\eta$ in the distance,
	\begin{equation}
		T_0 = C - \lambda^{-1}\ln\!\left(\frac{|\rho_{c,0}-\rho_{c,0}^{\star}|}{\rho_{c,0}^{\star}}+\eta\right) ,
		\label{eq:T0_noise}
	\end{equation}
	describes the saturation with $\eta\lesssim10^{-6}$. Since $\eta$ grows when
	the grid is coarsened, the floor is numerical. It limits how closely the
	threshold can be approached, but not the measured $\lambda$, which we fit only
	well above $\eta$ (Sec.~\ref{sec:SM_lifetimes}).
	
	\begin{figure}[t]
		\includegraphics[width=\columnwidth]{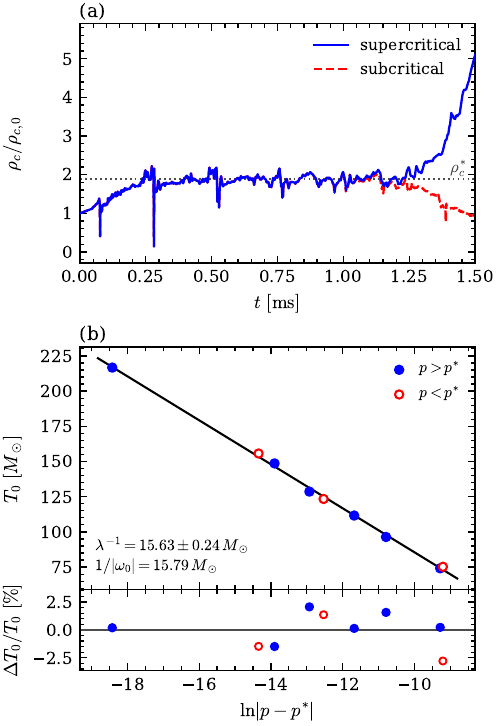}
		\caption{Velocity-induced critical collapse of a stable hybrid star ($M_B=2.471\,\Msun$).
			(a) Normalized central density for the two runs bracketing the threshold, $v_0=0.05253389\,c$ (solid, collapse) and $0.05253388\,c$ (dashed, no collapse). The dotted line marks $\rho_c^{*}$ of the unstable hybrid star with the same $M_B$.
			(b) Lifetime $T_0$ against $\ln|v_0-v_0^{*}|$ for collapsing (filled) and non-collapsing (open) runs, with the fit of Eq.~\eqref{eq:T0_scaling}. The lower panel shows the relative residuals.}
		\label{fig:crit}
	\end{figure}
	
	\textit{Velocity-induced collapse.}---Conversion acts on the star like an
	internal inward kick. As a control, we drive a stable hybrid star of the same
	EoS ($M=2.099\,\Msun$, $M_B=2.471\,\Msun$) with the inward velocity profile
	$V(r)=(v_0/2)(x^3-3x)$, $x=r/R_*$, and bisect in $v_0$ to $10^{-8}c$
	($v_0^{*}\simeq0.0525\,c$). Near threshold the star again lingers at the
	unstable hybrid star with its own baryon mass
	($\rho_c^{*}\simeq2.61\times10^{15}\gcc$, $R=10.13$~km) before collapsing or
	rebounding [Fig.~\ref{fig:crit}(a)], and the lifetimes on both sides follow
	Eq.~\eqref{eq:T0_scaling} with $\lambda^{-1}=15.6\pm0.2\pm0.3\,\Msun$, compared
	with $1/|\omega_0|=15.8\,\Msun$ for that star [Fig.~\ref{fig:crit}(b)].
	Both channels thus select the unstable hybrid star of equal baryon mass as the
	critical solution, with the lifetime set by its fundamental radial mode. A kick
	can in principle bring any stable hybrid star to threshold, whereas conversion
	of a given hadronic sequence selects a single critical star. Unlike the
	conversion family, the velocity family shows no floor down to
	$|v_0-v_0^{*}|=10^{-8}c$, which locates the origin of the floor at the onset of
	conversion. 
	
	\textit{Finite conversion speed.}---Instantaneous conversion is an
	idealization. We therefore repeat the bisection for the
	$\Delta\rho/\rho_t=0.15$ EoS with two finite-rate prescriptions (Sec.~\ref{sec:SM_conversion} of the Supplemental Material). Above
	the onset density, the cold pressure and energy density are interpolated
	between the hadronic and hybrid EoSs through a converted fraction $f(r,t)$,
	which either grows uniformly, $f=1-e^{-t/\tau}$, or is set by a front moving
	outward from the center at speed $v_f$; the energy released by conversion is
	deposited as heat. For $\tau=10^{-6}$~s and $v_f\ge0.95\,c$ the evolution of
	$\rho_c$ is indistinguishable from the instantaneous case, and for
	$\tau\le10^{-5}$~s and $v_f\ge0.5\,c$ the window stays within $5\%$ of its
	instantaneous width (Table~\ref{tab:finite_conversion}). Slower conversion
	weakens the initial pressure drop~\cite{Pereira2018}, so the core falls in less
	violently, and the window narrows: it retains about $70\%$ of its width for
	$\tau=5\times10^{-5}$~s and $v_f=0.1\,c$, a third for $\tau=10^{-4}$~s, and a
	quarter for $v_f=0.05\,c$. The window thus closes once conversion takes about
	as long as the dynamical time of the core, $\sim0.1$~ms. Throughout, the
	threshold remains a Type-I critical surface.
	
	\begin{table}[t]
		\caption{Width of the collapse window below the static threshold for the
			$\Delta\rho/\rho_t=0.15$ EoS of Table~\ref{tab:thresholds}, in baryon mass
			($\Delta M_B$) and gravitational mass ($\Delta M$, defined as in
			Table~\ref{tab:thresholds}), for instantaneous conversion and the two
			finite-rate models: homogeneous relaxation with e-folding time $\tau$, and an
			outward-propagating front with speed $v_f$. The last column gives $\Delta M_B$
			relative to its instantaneous value. Masses in $M_\odot$.}
		\label{tab:finite_conversion}
		\begin{ruledtabular}
			\begin{tabular}{llccc}
				Model & Parameter & $\Delta M_B$ & $\Delta M$ & Fraction \\
				\hline
				Instantaneous & --                                  & 0.0748  & 0.0512  & 1    \\
				\hline
				Homogeneous & $\tau = 1\times10^{-5}\,\mathrm{s}$       & 0.07172 & 0.04895 & 0.96 \\
				& $\tau = 5\times10^{-5}\,\mathrm{s}$       & 0.05248 & 0.03469 & 0.70 \\
				& $\tau = 1\times10^{-4}\,\mathrm{s}$       & 0.02478 & 0.01421 & 0.33 \\
				\hline
				Front   & $v_f = 0.90\,c$ & 0.07162 & 0.04888 & 0.96 \\
				& $v_f = 0.75\,c$ & 0.07151 & 0.04880 & 0.96 \\
				& $v_f = 0.50\,c$ & 0.07143 & 0.04874 & 0.95 \\
				& $v_f = 0.10\,c$ & 0.05380 & 0.03566 & 0.72 \\
				& $v_f = 0.05\,c$ & 0.01850 & 0.00958 & 0.25 \\
			\end{tabular}
		\end{ruledtabular}
	\end{table}


	\textit{Discussion.}---A first-order PT can serve as a natural trigger that drives a NS across the migration-collapse threshold. Conversion launches an inward flow that overshoots that equilibrium, and stars up to $0.117\,\Msun$ in baryon mass below the static threshold collapse, most of them to BHs lighter than the maximum hybrid mass. The dynamical threshold is a Type-I critical point. Its attractor is the unstable hybrid star with the same $M_B$, and the time spent near it is controlled by that star's fundamental radial mode. Stable hybrid stars driven by velocity perturbations are drawn to the same kind of attractor. In simulations, the onset of a strong transition also sets a resolution-dependent floor on how finely the threshold can be tuned. The measured growth rate remains reliable for $|\rho_{c,0}-\rho_{c,0}^{\star}|/\rho_{c,0}^{\star}\gg\eta$.

    For this channel to operate in nature, a hadronic star must exceed the onset density while its conversion is delayed, e.g.\ by quark-matter nucleation~\cite{Bombaci2004}, and must then convert rapidly, within $\lesssim0.1$~ms. Accretion, spin-down of a supramassive remnant, or proto-NS cooling can bring it there quasi-statically: the slow process sets the trigger, and the conversion sets the dynamics. The metastable hadronic stars must coexist with hybrid stars of the same baryon mass, as in the two-family scenario~\cite{Drago2014}.
	
	The PT is expected to release $\Delta M\,c^2\approx10^{52}$\,erg within $\lesssim1\ms$, but if a BH is formed almost all of it is plunges, thus having no transient signature. A magnetized star, however, loses its magnetosphere when the horizon forms, $0.3$--$1.3\ms$ after conversion. This could power a millisecond radio burst as in the ``blitzar'' scenario~\cite{Falcke2014,Most2018}. Additionally, in an ideal scenario, a non-spherical collapse would end in a BH ringdown at $f\approx5.6$--$5.8$~kHz with a damping time of $\approx0.12\ms$ for $M_{\rm BH}=2.1$--$2.25\,\Msun$~~\cite{Berti2009}.
	
	Thus, the natural extension of the work would be to run multidimensional simulations for rotating stars in which deflagration fronts can become unstable and turbulent, along with a broader survey of $\Delta\rho/\rho_t$ and onset density, to establish the width of the band in natural scenarios. Within these limits, our results identify a channel for BH formation below the static threshold, including BHs lighter than the most massive hybrid star, in NSs that undergo a strong first-order transition.

\begin{acknowledgments}
The authors thank the Indian Institute of Science Education and Research (IISER) Bhopal for providing all the research and infrastructure facilities. We thank Asim Kumar Saha for providing the equation of state used in this work, and E.~O'Connor and C.~D.~Ott for making \GRoneD\ publicly available. We used Claude Opus~5.5 (Anthropic) to assist with developing the manuscript. 
\end{acknowledgments}

    \textit{Data and code availability.}---The piecewise-polytrope parameters, mass–radius sequences, and all data presented in this Letter are provided with the article. Raw simulation outputs are available from the authors upon reasonable request.

	\bibliography{references}
	
	\clearpage
\setcounter{secnumdepth}{3}
\setcounter{section}{0}
\setcounter{equation}{0}
\setcounter{figure}{0}
\setcounter{table}{0}
\renewcommand{\thesection}{S\arabic{section}}
\renewcommand{\theequation}{S\arabic{equation}}
\renewcommand{\thefigure}{S\arabic{figure}}
\renewcommand{\thetable}{S\arabic{table}}
\renewcommand{\theHsection}{S\arabic{section}}
\renewcommand{\theHequation}{S\arabic{equation}}
\renewcommand{\theHfigure}{S\arabic{figure}}
\renewcommand{\theHtable}{S\arabic{table}}

\begin{center}
	\textbf{\large Supplemental Material}
\end{center}

\section{Numerical framework}
\label{sec:SM_numerics}

\textit{Field equations.}---\GRoneD~\cite{OConnor2010} evolves the
Einstein--Euler system in spherical symmetry in radial gauge and polar slicing,
\begin{equation}
	ds^2=-\alpha^2dt^2+X^2dr^2+r^2d\Omega^2,\qquad X=\Big(1-\frac{2m}{r}\Big)^{-1/2},
\end{equation}
with lapse $\alpha$ and enclosed gravitational mass $m(r)$ ($G=c=\Msun=1$
unless stated otherwise). The Hamiltonian constraint and the slicing condition
read
\begin{align}
	\partial_r m&=4\pi r^2\big(\rho hW^2-P\big),\\
	\partial_r\ln\alpha&=X^2\Big[\frac{m}{r^2}+4\pi r\big(\rho hW^2v^2+P\big)\Big],
\end{align}
where $v$ is the Eulerian velocity, $W=(1-v^2)^{-1/2}$, and
$h=1+\epsilon+P/\rho$ is the specific enthalpy, with $\epsilon$ the specific
internal energy. The conserved variables are $D=X\rho W$, $S^r=\rho hW^2v$, and
$E=\rho hW^2-P-D$. Fluxes are computed with the HLLE solver~\cite{Harten1983}
and second-order MC-limited reconstruction~\cite{vanLeer1977}, and the time
integration uses a second-order Runge--Kutta scheme with Courant factor $0.5$.
The grid is uniform with $\Delta r=20$~m out to $20$~km and geometrically
stretched beyond, up to $500$~km.

\textit{Code extensions.}---We extended the public code as follows; the patch is
part of the data package.
(i) A piecewise-polytropic cold EoS with a thermal component,
\begin{equation}
	P=P_{\rm cold}(\rho)+(\Gamma_{\rm th}-1)\rho\,[\epsilon-\epsilon_{\rm cold}(\rho)],
	\qquad\Gamma_{\rm th}=1.75,
	\label{eq:thermal}
\end{equation}
with the sound speed floored at $10^{-3}P/\rho$ as a safeguard in segments with
very small $c_s^2$.
(ii) TOV initial data on the cold EoS.
(iii) Radial velocity perturbations, imposed iteratively together with the
constraint equations so that the initial data satisfy the
constraints~\cite{Gourgoulhon1992}.
(iv) A robust atmosphere: a zone is reset to the floor when
$D/X<10\rho_{\rm floor}$, and a failed primitive-variable inversion below
$10^{11}\gcc$ resets the zone instead of stopping the run. The original
``freeze at atmosphere density'' rule is disabled, because it removed mass from
the stellar surface.
(v) A guard on the CFL condition against unphysical sound speeds in surface
zones.
(vi) Diagnostics: central density and lapse, minimum lapse, total gravitational
and baryon mass, and the masses in the quark and mixed phases.
(vii) Finite-rate conversion (Sec.~\ref{sec:SM_conversion}).

\textit{Collapse criterion.}---A run is classified as collapsing once the
central lapse falls below $\alpha_c=0.05$, beyond which the central density
grows without bound; representative runs evolved further approach horizon
formation. Runs that do not collapse are evolved for $5$~ms.

\textit{Linear radial modes.}---Frequencies and growth rates of the radial modes
are obtained from the self-adjoint form of the radial-perturbation
equation~\cite{chandrasekhar1964,Gondek1997,Kokkotas2001}, discretized with
linear finite elements in $\xi=\delta r/r$. A lumped mass matrix turns the
problem into a symmetric tridiagonal eigenvalue problem, which yields all modes
in order, including those with $\omega^2<0$; the growth rate of an unstable
mode is $|\omega|=\sqrt{-\omega^2}$. Nodes are placed at every break of the
piecewise polytrope, the result is Richardson-extrapolated from $N$ and $2N$
elements, and the adiabatic index is the equilibrium one,
$\Gamma=(\varepsilon+P)c_s^2/P$, so the mixed phase is assumed to stay in local
equilibrium, as in the evolutions. The solver reproduces the change of sign of
$\omega_0^2$ at the maximum of the $M(\varepsilon_c)$ curve of a $\Gamma=2$
polytrope to $0.02\%$. Here $\omega$ refers to Schwarzschild coordinate time,
the same time as that of \GRoneD, so it can be compared with the measured
lifetimes without a lapse rescaling.

\section{Equation of state}
\label{sec:SM_eos}

\textit{Table.}---The EoS~\cite{Saha:2025} is tabulated in $\varepsilon$ and
$P$ [MeV\,fm$^{-3}$], $n_B$ [fm$^{-3}$], and chemical potential. The rest-mass
density is $\rho=m_bn_B$ with $m_b=1.674927\times10^{-24}$~g. A single
constant-pressure step marks the Maxwell transition, at
$P_t=99.91\,{\rm MeV\,fm^{-3}}$ from $\varepsilon_t=425.4$ to
$\varepsilon_q=645.7\,{\rm MeV\,fm^{-3}}$, i.e.\ from $\rho_t=7.052\times10^{14}$
to $\rho_q=1.001\times10^{15}\gcc$. The jump $\Delta\varepsilon/\varepsilon_t=0.518$
lies below the Seidov limit $0.852$~\cite{Seidov1971}.

\textit{Piecewise-polytropic fit.}---In segment $i$ the pressure is
$P=K_i\rho^{\Gamma_i}$ for $\rho_{i-1}\le\rho<\rho_i$~\cite{Read2009}, and the
energy density follows from the first law,
\begin{equation}
	\varepsilon=(1+a_i)\rho+\frac{K_i}{\Gamma_i-1}\rho^{\Gamma_i},
\end{equation}
with $K_i$ and $a_i$ fixed by continuity of $P$ and $\varepsilon$. The fit is
done in $x=\log_{10}\rho$, $y=\log_{10}P$ by greedy segmentation: a segment is
extended one tabulated point at a time while the rms residual of its linear
least-squares fit stays below $0.01$~dex, with a minimum of three points, and
its slope gives $\Gamma_i$. The hadronic part ($10^{14.3}\gcc\le\rho\le\rho_t$)
and the quark part ($\rho\ge\rho_q$) are fitted separately. Below the first
fitted segment we use the four-piece SLy crust~\cite{Douchin2001,Read2009},
joined at $\log_{10}\rho=14.058$.

	\begin{figure}[t]
		\includegraphics[width=\columnwidth]{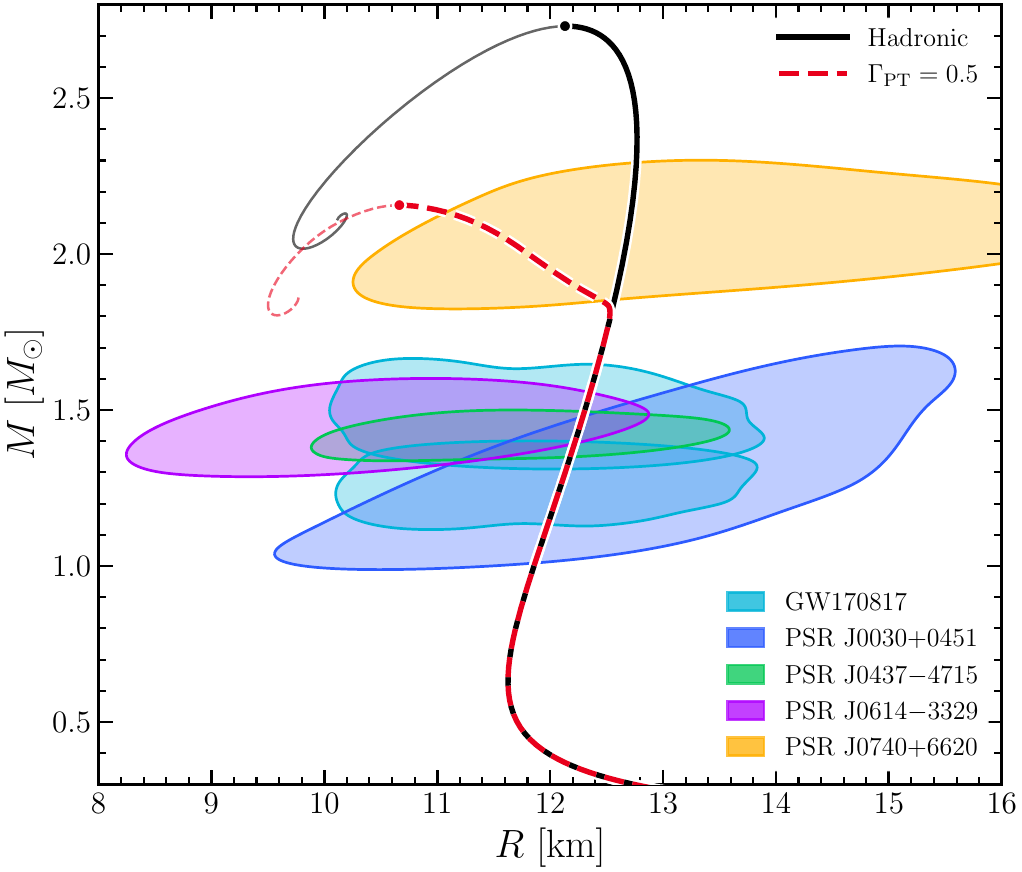}
		\caption{Mass--radius sequences for the hadronic EoS and the EoS with a phase transition. Translucent and dashed curves denote unstable configurations. Observational constraints are shown for comparison.}
		\label{fig:eos}
	\end{figure}

\textit{Causality.}---Without a further constraint, the fit absorbs the steep
rise of the pressure just above $\rho_q$ into one segment that reaches
$c_s^2=1.11$. We therefore also stop a segment once
\begin{equation}
	c_s^2=\frac{\Gamma_iP}{\varepsilon+P}>0.95
\end{equation}
at its last point, with $P$ and $\varepsilon$ from the table. Because this
criterion uses the tabulated values, the fitted EoS can exceed it slightly: its
largest sound speed, $c_s^2=0.997$, is reached at the end of the first quark
segment, and the fit is causal everywhere. It reproduces the tabulated pressure
to $0.012$~dex rms ($0.027$~dex maximum) and the energy density to $0.96\%$ rms.

\textit{Mixed phase.}---At the Maxwell plateau $c_s^2=0$, which makes the
hydrodynamic system only weakly hyperbolic. For the evolutions the plateau is
therefore replaced by a segment with $\Gamma_{\rm mix}=0.5$, which joins the
hadronic fit at $\log_{10}\rho=14.8308$ and the quark fit at $\rho_q$. The
resulting EoS (Table~\ref{tab:eos_pp_main}) has three hadronic segments, the
mixed phase, and six quark segments.

\begin{table}[t]
	\caption{Piecewise-polytropic parametrization of the hybrid EoS of the Letter,
		in the convention of Table~\ref{tab:eos_pp}: in segment $i$,
		$P=K_i\rho^{\Gamma_i}$ for $\rho_{i-1}\le\rho<\rho_i$ (rest-mass density in
		g\,cm$^{-3}$), with the $K_i$ fixed by continuity of the pressure from the SLy
		crust at $\log_{10}\rho=14.0582$. Segments 0--2 coincide with the hadronic EoS,
		segment~3 is the mixed phase with $\Gamma_{\rm mix}=0.5$ (bold), of width
		$\Delta\rho/\rho=0.478$, and segments 4--9 describe quark matter.}
	\label{tab:eos_pp_main}
	\begin{ruledtabular}
		\begin{tabular}{rcc}
			$i$ & $\log_{10}\rho_{i-1}$ & $\Gamma_i$ \\
			\hline
			0 & $-$ & 2.266487 \\
			1 & 14.5360543 & 3.984579 \\
			2 & 14.6590154 & 5.350409 \\
			3 & 14.8307790 & \textbf{0.500000} \\
			4 & 15.0004089 & 6.069784 \\
			5 & 15.0212381 & 5.027293 \\
			6 & 15.0415510 & 4.218746 \\
			7 & 15.0711280 & 2.841472 \\
			8 & 15.2246400 & 1.455011 \\
			9 & 15.4003572 & 0.939793 \\
		\end{tabular}
	\end{ruledtabular}
\end{table}

\section{Dependence on the strength of the phase transition}
\label{sec:Strength_PT}

\begin{figure}[t]
	\includegraphics[width=\columnwidth]{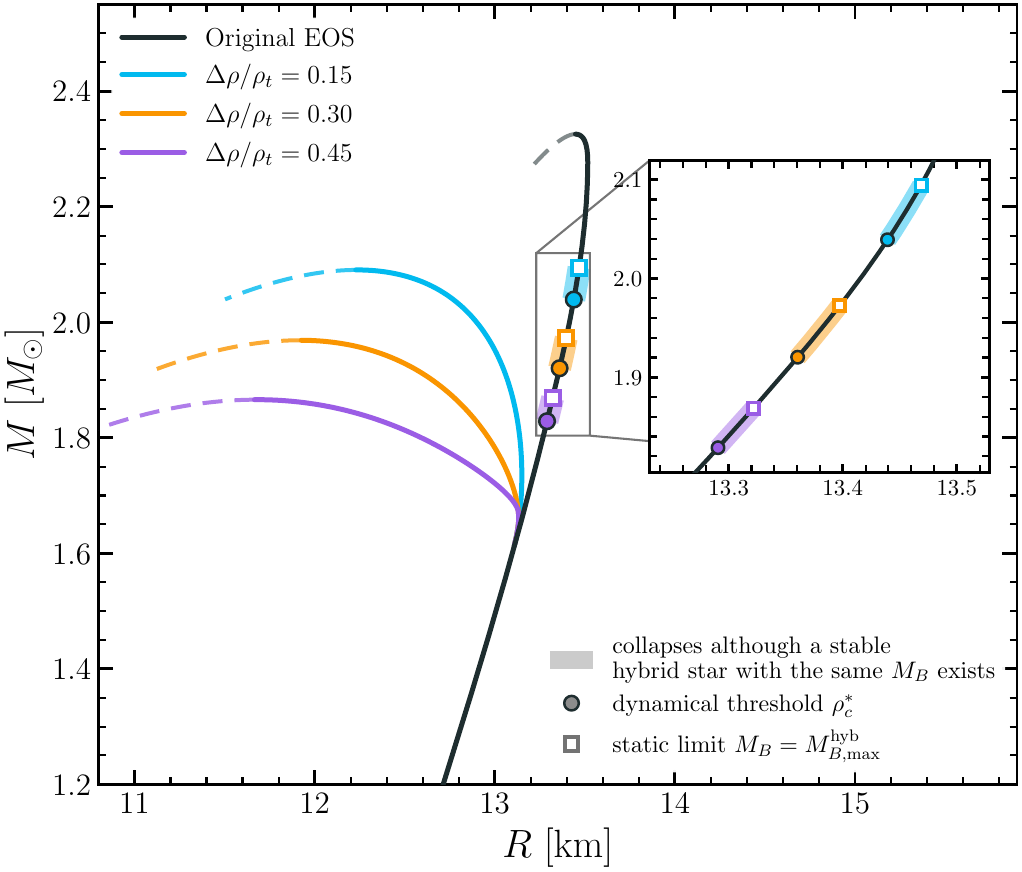}
	\caption{Mass--radius relations of the hadronic EoS (dark) and of the hybrid
		EoSs with $\Delta\rho/\rho_t=0.15$, $0.30$, and $0.45$ (coloured); dashed lines
		mark unstable configurations. On the hadronic branch, the shaded segment of each
		colour contains the stars that collapse on conversion although a stable hybrid
		star with the same baryon mass exists. It extends from the dynamical threshold
		$\rho_{c,0}^{\star}$ found by bisection (filled circle) to the static limit
		$M_B=M_{B,\max}^{\rm hyb}$ (open square). The inset enlarges this part of the
		hadronic branch.}
	\label{fig:3eos}
\end{figure}

The window of hadronic stars that collapse on conversion, although a stable
hybrid star with the same baryon mass exists, could in principle be a feature of
the particular EoS of the Letter. We therefore repeat the analysis for a
different hadronic EoS combined with phase transitions of increasing strength
(Table~\ref{tab:eos_pp}). All of them set in at the same density,
$\rho_t=10^{14.747122}\gcc$, where the hadronic star has $M=1.6\,\Msun$.
Between $\rho_t$ and $\rho_t+\Delta\rho$ the matter is described by a soft
segment with $\Gamma_{\rm PT}=0.5$, and above it by quark matter with constant
sound speed, $c_s^2=1/3$. We consider $\Delta\rho/\rho_t=0.15$, $0.30$, and
$0.45$; a larger width corresponds to a stronger transition and a lower maximum
mass of the hybrid branch (Fig.~\ref{fig:3eos}).

\begin{table*}[t]
	\caption{Piecewise-polytropic parametrization of the hadronic EoS and of the
		three hybrid EoSs of Sec.~\ref{sec:Strength_PT}. In segment $i$ the pressure is
		$P=K_i\rho^{\Gamma_i}$ for $\rho_{i-1}\le\rho<\rho_i$, with $\rho$ the rest-mass
		density in g\,cm$^{-3}$; the table lists the lower dividing density
		$\log_{10}\rho_{i-1}$ and the adiabatic index $\Gamma_i$. The constants $K_i$
		follow from continuity of the pressure, with $\log_{10}P=33.649778$ (cgs) at
		$\log_{10}\rho=14.4695$. Below segment~0 we use the four-piece SLy crust of
		Ref.~\cite{Read2009}, joined where the pressures match
		($\log_{10}\rho\simeq14.047$), and the last segment extends to arbitrarily high
		density. All four EoSs share segments 0--2 and coincide below the transition
		density $\log_{10}\rho_t=14.747122$ ($P_t=53.751$\,MeV\,fm$^{-3}$). In the
		hybrid EoSs the transition is a soft segment with $\Gamma_{\rm PT}=0.5$ (bold)
		of width $\Delta\rho/\rho_t$, followed by quark matter with $c_s^2=1/3$,
		represented by segments whose $\Gamma_i$ approach $4/3$; the corresponding bag
		constants are $B^{1/4}=142.5$, $149.8$, and $156.3$\,MeV.}
	\label{tab:eos_pp}
	\begin{ruledtabular}
		\begin{tabular}{r cc cc cc cc}
			& \multicolumn{2}{c}{Hadronic} & \multicolumn{2}{c}{$\Delta\rho/\rho_t = 0.15$} & \multicolumn{2}{c}{$\Delta\rho/\rho_t = 0.30$} & \multicolumn{2}{c}{$\Delta\rho/\rho_t = 0.45$} \\
			$i$ & $\log_{10}\rho_{i-1}$ & $\Gamma_i$ & $\log_{10}\rho_{i-1}$ & $\Gamma_i$ & $\log_{10}\rho_{i-1}$ & $\Gamma_i$ & $\log_{10}\rho_{i-1}$ & $\Gamma_i$ \\
			\hline
			0 & $-$ & 2.444947 & $-$ & 2.444947 & $-$ & 2.444947 & $-$ & 2.444947 \\
			1 & 14.469500 & 3.596919 & 14.469500 & 3.596919 & 14.469500 & 3.596919 & 14.469500 & 3.596919 \\
			2 & 14.602600 & 5.572017 & 14.602600 & 5.572017 & 14.602600 & 5.572017 & 14.602600 & 5.572017 \\
			3 & 14.773300 & 2.577157 & 14.747122 & \textbf{0.500000} & 14.747122 & \textbf{0.500000} & 14.747122 & \textbf{0.500000} \\
			4 & 14.851700 & 0.487774 & 14.807820 & 2.508524 & 14.861065 & 2.680601 & 14.908490 & 2.840725 \\
			5 & 15.262000 & 1.098243 & 14.829100 & 2.373068 & 14.881294 & 2.521546 & 14.928256 & 2.659325 \\
			6 & 15.410400 & 4.891912 & 14.851444 & 2.252649 & 14.903079 & 2.379682 & 14.949036 & 2.500281 \\
			7 & 15.484268 & 4.442932 & 14.875384 & 2.144552 & 14.926420 & 2.254714 & 14.971336 & 2.359206 \\
			8 & 15.501000 & 0.968738 & 14.900920 & 2.047822 & 14.950798 & 2.144605 & 14.995157 & 2.234292 \\
			9 & 15.711200 & 2.186743 & 14.928052 & 1.962213 & 14.976732 & 2.046590 & 15.020499 & 2.123813 \\
			10 & 15.931600 & 0.674532 & 14.956248 & 1.886459 & 15.004223 & 1.959482 & 15.047361 & 2.026173 \\
			11 & 16.065600 & 1.344979 & 14.986040 & 1.818318 & 15.033269 & 1.881568 & 15.075743 & 1.939923 \\
			12 &  &  & 15.017961 & 1.729161 & 15.064391 & 1.812077 & 15.105646 & 1.863204 \\
			13 &  &  & 15.087121 & 1.631865 & 15.097068 & 1.750308 & 15.137576 & 1.794630 \\
			14 &  &  & 15.164261 & 1.554606 & 15.131820 & 1.670312 & 15.171534 & 1.733597 \\
			15 &  &  & 15.250978 & 1.493578 & 15.207030 & 1.583218 & 15.207519 & 1.679482 \\
			16 &  &  & 15.348866 & 1.447596 & 15.292095 & 1.514504 & 15.245531 & 1.609610 \\
			17 &  &  & 15.451010 & 1.413168 & 15.388052 & 1.462753 & 15.328651 & 1.534010 \\
			18 &  &  & 15.567519 & 1.386790 & 15.488677 & 1.423971 & 15.422921 & 1.476888 \\
			19 &  &  & 15.701052 & 1.367188 & 15.603307 & 1.394274 & 15.522259 & 1.434134 \\
			20 &  &  & 15.857460 & 1.348313 & 15.734535 & 1.372164 & 15.634775 & 1.401397 \\
			21 &  &  & 16.258058 & 1.336506 & 15.888067 & 1.356534 & 15.764016 & 1.376926 \\
			22 &  &  &  &  & 16.064420 & 1.346018 & 15.915051 & 1.359584 \\
			23 &  &  &  &  & 16.280194 & 1.337162 & 16.087880 & 1.347836 \\
			24 &  &  &  &  &  &  & 16.299734 & 1.337828 \\
		\end{tabular}
	\end{ruledtabular}
\end{table*}

For each hybrid EoS we evolve unperturbed hadronic stars with the hybrid EoS and
locate the dynamical threshold by bisection in the initial central density
$\rho_{c,0}$, with the same grid, atmosphere, and collapse criterion as in the
Letter, until the bracket is narrower than $10^{-9}\rho_{c,0}$. We compare the
threshold $\rho_{c,0}^{\star}$ with the static limit, i.e.\ the hadronic star
whose baryon mass equals the maximum baryon mass of the stable hybrid branch.
Hadronic stars beyond the static limit have no hybrid counterpart and must
collapse; stars between $\rho_{c,0}^{\star}$ and the static limit have one, but
collapse nevertheless.

The thresholds lie at $M_*=2.0393$, $1.9203$, and $1.8286\,\Msun$ for
$\Delta\rho/\rho_t=0.15$, $0.30$, and $0.45$, compared with hybrid maximum
masses of $2.0905$, $1.9688$, and $1.8659\,\Msun$ and maximum baryon masses of
$2.4271$, $2.2630$, and $2.1262\,\Msun$ (Table~\ref{tab:thresholds}). For every
transition strength the threshold lies below the static limit, so a window of
statically stable but dynamically collapsing stars is present in every case
(shaded segments in Fig.~\ref{fig:3eos}), and it is not an artifact of the EoS
of the Letter. Its width decreases with increasing $\Delta\rho/\rho_t$.

\section{Lifetime extraction and critical scaling}
\label{sec:SM_lifetimes}

\textit{Families and bisection.}---We study two one-parameter families of
initial data. In the \emph{conversion family}, metastable hadronic stars are
evolved with the hybrid EoS without perturbation, and the parameter is the
initial (TOV) central density, $p=\rho_{c,0}$. In the \emph{velocity family}, a
representative stable hybrid star ($M=2.0994\,\Msun$, $M_B=2.471\,\Msun$)
receives the inward profile $V(r)=(v_0/2)(x^3-3x)$, $x=r/R_*$, and the
parameter is $p=v_0$. We bisect each family until the bracketing pair
$p_{\rm lo}$ (no collapse) and $p_{\rm hi}$ (collapse) differs by
$5.3\times10^{-10}$ in relative terms (conversion family, 28 rounds) or by
$10^{-8}c$ (velocity family), and identify $p^{*}$ with $p_{\rm lo}$; the
difference is negligible compared with all distances used in the fits. For the
conversion family we use the dimensionless distance
$|\rho_{c,0}-\rho_{c,0}^{\star}|/\rho_{c,0}^{\star}$, which changes only the
constant $C$ in Eq.~\eqref{eq:T0_scaling}. The bisection leaves only three non-collapsing runs in the scaling range; the
others lie either within $2\times10^{-6}$ of the threshold, inside the noise
floor, or beyond $2\times10^{-3}$. We therefore add non-collapsing runs at
relative offsets $-10^{-5}$, $-3\times10^{-5}$, $-10^{-4}$, $-3\times10^{-4}$,
and $-10^{-3}$, so that both branches span a comparable range in distance.

\textit{Observable.}---Following Ref.~\cite{Noble2016}, we define the lifetime
$T_0(p)$ relative to the marginally subcritical run $p_{\rm lo}$, the evolution
that stays near the critical solution longest. The observable is the central
rest-mass density $\rho_c(t)$, written at every time step. Times are \GRoneD\
coordinate times, equal to the proper time of a static observer at infinity,
and are quoted in units of $\Msun$ ($1\,{\rm ms}=203.03\,\Msun$). Where
restarts leave repeated output times, we keep the last value written.

During the hover, $\rho_c$ is modulated by up to $\pm25\%$ (rms $\approx10\%$)
on time scales of $0.1$--$0.2$~ms.

\textit{Smoothed deviation.}---Because the central density exhibits pronounced single-peak spikes and modulations near the critical region, extracting the lifetime of the configuration in this regime is not straightforward. So, we interpolate $\rho_c(t)$ onto a uniform grid
($\Delta t=0.1\,\Msun$) and average it over a centred window of width
$W=30\,\Msun\approx0.15$~ms, comparable to one pulsation period:
\begin{equation}
	\langle\rho_c\rangle_W(t)=\frac{1}{W}\int_{t-W/2}^{t+W/2}\rho_c(t')\,dt' .
\end{equation}
The lifetime is defined from
\begin{equation}
	\delta(t;p)=\left|\frac{\langle\rho_c\rangle_W(t;p)}{\langle\rho_c\rangle_W(t;p_{\rm lo})}-1\right| .
	\label{eq:S_delta}
\end{equation}
In the linear regime $\delta(t;p)\propto|p-p^{*}|\,e^{\lambda t}$, so the
curves $\delta/|p-p^{*}|$ of different runs must coincide. They do, for $75\lesssim T_0/\Msun\lesssim215$ in the velocity family and
$70\lesssim T_0/\Msun\lesssim155$ in the conversion family, except for runs
close to the noise floor.
\textit{Permanent departure.}---$T_0(p)$ is the time of the permanent departure
of $\delta$ above a threshold $\epsilon$: the last instant at which $\delta$
crosses $\epsilon$ from below, obtained by linear interpolation. Transient
excursions during the initial adjustment are thereby ignored. The comparison is
restricted to times before the reference run itself leaves the critical
solution, i.e.\ before $p_{\rm lo}$ and $p_{\rm hi}$ separate at the same
$\epsilon$; in the conversion family we extend this window by $30\,\Msun$, so
that runs near the noise floor are not truncated. A run that never departs
permanently within the window, or that starts above $\epsilon$, yields no
lifetime. We use $\epsilon=0.005$, $0.01$, and $0.02$ for the velocity family
and, because nearby runs already differ by $\sim10^{-2}$ during the hover,
$\epsilon=0.02$, $0.05$, and $0.10$ for the conversion family.

\textit{Fits.}---We fit Eq.~\eqref{eq:T0_scaling} by ordinary least squares,
with the statistical error of the slope estimated from the scatter of the
residuals, using (i) a single line through both branches, which gives the
quoted $\lambda^{-1}$, and (ii) a common slope with separate intercepts $C_\pm$
for collapsing and non-collapsing runs, which tests for an offset
$\Delta C=C_--C_+$. The velocity fit uses nine runs with
$10^{-8}c\le|v_0-v_0^{*}|\lesssim10^{-4}c$; runs with
$|v_0-v_0^{*}|\gtrsim10^{-3}c$ have lifetimes comparable to the initial
transient. The conversion fit uses fifteen runs (eight non-collapsing, seven
collapsing) with $5\times10^{-6}\le|\rho_{c,0}-\rho_{c,0}^{\star}|/\rho_{c,0}^{\star}\le1.5\times10^{-3}$;
the upper limit excludes pre-asymptotic runs, and the lower limit lies a few
times above the noise floor. Since the choice of $\epsilon$ is arbitrary, we
quote for each family the mean slope over the three thresholds, with the mean
fit error as statistical and half the spread between thresholds as systematic
error. As an extraction-free cross-check we also fit the collapse times
$t_{\rm col}$ of the supercritical runs, which equal $T_0$ plus the time from
departure to $\alpha_c=0.05$ and therefore have the same slope; two runs that
were classified with the earlier criterion $\alpha_c<0.08$ are left out of this
fit.

\begin{table}[t]
	\caption{Inverse growth rate $\lambda^{-1}$ (in $\Msun$) from single-line fits
		of Eq.~\eqref{eq:T0_scaling}, compared with $1/|\omega_0|$ of the corresponding
		critical star. Errors of single fits are statistical; $\epsilon$-averaged values
		give the mean statistical error and, as systematic error, half the spread
		between thresholds. Collapse times $t_{\rm col}$ require no lifetime
		extraction; at $20$~m, two runs whose collapse was recorded with a different
		lapse criterion are excluded.}
	\label{tab:S_lambda}
	\begin{ruledtabular}
		\begin{tabular}{llc}
			Family / variant & Setting & $\lambda^{-1}$ \\
			\hline
			Velocity, $\Delta r=20$~m & $\epsilon=0.005$ & $15.91\pm0.53$ \\
			& $\epsilon=0.01$ & $15.40\pm0.34$ \\
			& $\epsilon=0.02$ & $15.62\pm0.24$ \\
			& $\epsilon=0.02$, $W=20\,\Msun$ & $15.75\pm0.31$ \\
			& $\epsilon=0.02$, $W=45\,\Msun$ & $15.75\pm0.19$ \\
			& $\epsilon$-averaged & $15.6\pm0.4\pm0.3$ \\
			& $1/|\omega_0|$ & $15.79$ \\
			\hline
			Conversion, $\Delta r=20$~m & $\epsilon=0.02$ & $14.89\pm0.35$ \\
			& $\epsilon=0.05$ & $15.46\pm0.34$ \\
			& $\epsilon=0.10$ & $15.04\pm0.38$ \\
			& $\epsilon$-averaged & $15.1\pm0.4\pm0.3$ \\
			& collapse times $t_{\rm col}$ & $14.3\pm0.1$ \\
			Conversion, $\Delta r=40$~m & $\epsilon$-averaged & $14.4\pm0.9\pm0.3$ \\
			& collapse times $t_{\rm col}$ & $13.6\pm0.6$ \\
			Conversion & $1/|\omega_0|$ & $14.60$ \\
		\end{tabular}
	\end{ruledtabular}
\end{table}

\textit{Results.}---Table~\ref{tab:S_lambda} summarizes the fits. In the
velocity family the two-branch fit gives an intercept difference consistent
with zero ($\Delta C=1.4\pm1.4\,\Msun$ at $\epsilon=0.02$). In the conversion
family it gives $\Delta C=2.8\pm1.1$, $0.2\pm1.4$, and $3.9\pm1.0\,\Msun$ for the
three thresholds, without a systematic trend and small compared with
$T_0\approx70$--$160\,\Msun$, so both branches are described by a single line.
The rms residuals of the single-line fits are $1.7\,\Msun$ (velocity family,
$\epsilon=0.02$) and $2.1$--$2.4\,\Msun$ (conversion family), and all relative
residuals stay below $3.5\%$. The collapse times give
$\lambda^{-1}=14.3\pm0.1\,\Msun$ with residuals of $0.2\,\Msun$. At
$\Delta r=40$~m the lower end of the fit range lies only about twice above the
noise floor, which may bias the fitted $\lambda^{-1}$ low.

\textit{Noise floor.}---In the conversion family the lifetimes stop growing
below a distance of $\sim10^{-6}$ from the threshold [Fig.~\ref{fig:ptcrit}(b)
of the Letter], and the collapse times behave in the same way: the seven runs
closer than $1.5\times10^{-6}$ collapse after $212$--$272\,\Msun$ without any
trend with distance, whereas the scaling line would predict up to
$\approx330\,\Msun$. Nearby runs separate as the centre first enters the mixed
phase, much faster than the unstable mode can amplify their initial difference.
The onset of conversion thus gives every run a perturbation that is set by the
discretization rather than by its distance from the threshold.
Eq.~\eqref{eq:T0_noise} describes the resulting saturation through an effective
offset $\eta$. Its fitted value depends on the observable, $\sim7\times10^{-7}$
from $T_0$ and $\sim2\times10^{-7}$ from $t_{\rm col}$ at $\Delta r=20$~m, and
it grows to $\sim3\times10^{-6}$ at $40$~m, so the floor is numerical; we use
$\eta$ only to describe the floor, not as a measured quantity. Stable evolutions
at further resolutions ($30$ and $80$~m) could not be obtained, so the
convergence order is not determined. The velocity family, evolved with the same
EoS but without a conversion onset, shows no floor: its closest run, at
$|v_0-v_0^{*}|=10^{-8}c$, lies on the scaling line.

\textit{Critical stars.}---The critical stars are specified by their central
density $\rho_c^{*}$, the time-averaged plateau value of the near-critical
runs. For the conversion family, $\rho_c^{*}=(2.87\pm0.04)\times10^{15}\gcc$
gives $M=2.086\,\Msun$, $R=10.05$~km, $|\omega_0|/2\pi=2.214$~kHz, and a first
overtone at $4.40$~kHz. For the velocity family,
$\rho_c^{*}=2.61\times10^{15}\gcc$ gives $M=2.10\,\Msun$, $R=10.13$~km, and
$|\omega_0|/2\pi=2.046$~kHz. Because $\omega_0^2$ varies steeply with $\rho_c$
near the turning point, the uncertainty of $\rho_c^{*}$ propagates into
$1/|\omega_0|$.

\section{Finite conversion speed}
\label{sec:SM_conversion}

\textit{Model.}---In the instantaneous runs a hadronic TOV star is evolved with
the hybrid EoS from $t=0$. To relax this, we keep both cold EoSs and mix them
through a converted fraction $f(r,t)\in[0,1]$,
\begin{equation}
	\begin{aligned}
		P_{\rm cold} &= (1-f)\,P_{\rm had}(\rho)+f\,P_{\rm hyb}(\rho), \\
		\epsilon_{\rm cold} &= (1-f)\,\epsilon_{\rm had}(\rho)+f\,\epsilon_{\rm hyb}(\rho),
	\end{aligned}
	\label{eq:mix}
\end{equation}
with the thermal part of Eq.~\eqref{eq:thermal} on top. The two EoSs coincide
below the onset density, so $f$ acts only where conversion is possible. The
evolution carries $\rho$ and $\epsilon$, and conversion acts at fixed
$(\rho,\epsilon)$: it changes only the split of $\epsilon$ into cold and thermal
parts. The released energy, $\epsilon_{\rm had}-\epsilon_{\rm hyb}$, therefore
appears as heat without an additional source term, and its pressure,
$(\Gamma_{\rm th}-1)\rho(\epsilon_{\rm had}-\epsilon_{\rm hyb})$, partly offsets
the drop of the cold pressure. The fraction depends only on position and time
and is evaluated at the time of each Runge--Kutta substep. Conversion is local
and reversible: matter that decompresses below the onset density follows the
common EoS again.

We use two prescriptions, both starting at $t=0$ from the unperturbed hadronic
star:
\begin{align}
	\text{uniform:}\quad & f=1-e^{-t/\tau},\\
	\text{front:}\quad & f=\min\!\left[1,\max\!\left(0,\frac{v_ft-r}{w}\right)\right],
\end{align}
with a ramp of width $w=0.5$~km (25 zones) behind the front, so that even the
centre is fully converted only after $w/v_f\lesssim3\,\mu$s for $v_f\ge0.5c$.
Instantaneous conversion is recovered for $\tau\to0$, and for $v_f\to\infty$,
$w\to0$.

\textit{Procedure.}---For each $\tau$ and $v_f$ we bisect in $\rho_{c,0}$ for
the $\Delta\rho/\rho_t=0.15$ EoS of Sec.~\ref{sec:Strength_PT}, with the same
grid, atmosphere, and collapse criterion, and define the window as
$\Delta M_B=M_{B,\rm TOV}-M_{B,*}$, with $M_{B,\rm TOV}$ of the hybrid EoS.

\textit{Results.}---For $\tau = 10^{-6}\,\mathrm{s}$, the single-run evolution of the central density, $\rho_c$, closely reproduces that of the instantaneous case for the same initial stellar model. Similarly, for the front propagating at $v_f = 0.99c$ with a ramp width narrower than one grid zone, the evolution closely approaches the instantaneous case, with a small residual delay.
For $v_f\ge0.5c$ the window levels off at
$\Delta M_B\simeq0.0715\,\Msun$, and $\tau=10^{-5}$~s gives the same width,
$4$--$5\%$ below the instantaneous value (Table~\ref{tab:finite_conversion} of
the Letter). The window shrinks once the conversion time becomes comparable to
the dynamical time of the core, $\sim0.1$~ms: to $70\%$ and $33\%$ of its
instantaneous width for $\tau=5\times10^{-5}$ and $10^{-4}$~s, and to $72\%$ and
$25\%$ for $v_f=0.1c$ and $0.05c$. In all cases the near-threshold runs hover
near the unstable hybrid star of equal baryon mass, and their lifetimes follow
Eq.~\eqref{eq:T0_scaling}.


\end{document}